\documentclass[prb,twocolumn,superscriptaddress,showpacs]{revtex4-1}

\usepackage{graphicx}
\usepackage{dcolumn}
\usepackage{bm}
\usepackage{amsmath}
\usepackage{float}

\begin{document}

\newcommand{\SrAs}{SrFe$_{2}$As$_{2}$ }
\newcommand{\CaAs}{CaFe$_{2}$As$_{2}$ }
\newcommand{\BaAs}{BaFe$_{2}$As$_{2}$ }

\title{Itinerant Spin Excitations in SrFe$_{2}$As$_{2}$ Measured by Inelastic Neutron Scattering}

\author{R. A. Ewings}
\email{russell.ewings@stfc.ac.uk} \affiliation{ISIS Facility, STFC Rutherford Appleton Laboratory, Chilton, Didcot, Oxon, OX11 0QX, United Kingdom}
\author{T. G. Perring} \affiliation{ISIS Facility, STFC Rutherford Appleton Laboratory, Chilton, Didcot, Oxon, OX11 0QX, United Kingdom} \affiliation{Department of Physics and Astronomy, University College London, Gower Street, London WC1E 6BT, United Kingdom}
\author{J. Gillett} \affiliation{Cavendish Laboratory, University of Cambridge, JJ Thomson Avenue, Cambridge CB3 0HE, United Kingdom}
\author{S. D. Das} \affiliation{Cavendish Laboratory, University of Cambridge, JJ Thomson Avenue, Cambridge CB3 0HE, United Kingdom}
\author{S. E. Sebastian} \affiliation{Cavendish Laboratory, University of Cambridge, JJ Thomson Avenue, Cambridge CB3 0HE, United Kingdom}
\author{A. E. Taylor} \affiliation{ Department of Physics,
University of Oxford, Clarendon Laboratory, Parks Road, Oxford, OX1
3PU, United Kingdom}
\author{T. Guidi} \affiliation{ISIS Facility, STFC Rutherford Appleton Laboratory, Chilton, Didcot, Oxon, OX11 0QX, United Kingdom}
\author{A. T. Boothroyd} \affiliation{ Department of Physics,
University of Oxford, Clarendon Laboratory, Parks Road, Oxford, OX1
3PU, United Kingdom}

\date{\today}

\begin{abstract}
We report inelastic neutron scattering measurements of the magnetic excitations in SrFe$_{2}$As$_{2}$, the parent of a family of iron-based superconductors. The data extend throughout the Brillouin zone and up to energies of $\sim 260$\,meV. The spectrum calculated from a $J_1$--$J_2$ model does not accurately describe our data, and we show that some of the qualitative features the model fails to describe are readily explained by calculations from a 5-band itinerant mean-field model. In particular, the high-energy part of the spectra recorded above $T_{\rm{N}}$ do not differ significantly from those at low temperature, which is explained by the itinerant model and which has implications for theories of electronic nematic and orbital ordering.
\end{abstract}


\maketitle

\section{Introduction}

The new iron-based superconductors 
show an intriguing interplay between structure, magnetism, and superconductivity \cite{Norman-Physics-2008,Paglione-NatPhys-2010,Johnston-AdvPhys-2010}. It is likely that the paradigm to understand these materials will be rather different from that used to approach the cuprate superconductors, since the parent compounds for iron-based superconductors are bad metals, rather than Mott insulators. Theoretical estimates suggest that the electron--phonon interaction is not the primary component of the pairing interaction in the iron-based superconductors \cite{Boeri-condmat-2010}, and most attention is now on magnetic pairing mechanisms \cite{Mazin-Schmalian-review}. Much theoretical and experimental effort has therefore been devoted to understanding the magnetism in this family of materials \cite{Lumsden-JPCM-review}. A key issue is whether the magnetism is better described within a weak-coupling (itinerant) \cite{Singh-Du-2008} or strong-coupling (localized) \cite{Si localized PRL} picture.

One of the most direct ways to identify the character of the magnetism is to study the magnetic excitation spectrum by inelastic neutron scattering (INS). Previous INS measurements of the antiferromagnetic (AFM) 122-arsenides $X$Fe$_{2}$As$_{2}$ (where $X$ is an alkali-earth metal) have been interpreted in terms of linear spin-wave theory, predicated on local-moment $J_1$--$J_2$ models \cite{Our Ba122,Diallo Ca122,Pengcheng Ca122}. These models include nearest-neighbor ($J_{1}$) and in-plane diagonal next-nearest neighbor ($J_{2}$) Fe--Fe exchange interactions. A particularly interesting result from this approach when applied to \CaAs is the very large difference between $J_{1a}$ and $J_{1b}$, the two in-plane nearest-neighbor exchange parameters \cite{Pengcheng Ca122}. Various mechanisms have been proposed to explain this anisotropy, including electronic nematic ordering \cite{Fang-PRB77-Nematic}, orbital ordering \cite{Chen-PRB-OrbitalOrder}, and the crystal structure itself \cite{Park-PRB-2010}. An important piece of information, lacking up to now, is whether the anisotropy is modified on warming above the combined magnetic and structural transition temperature ($T_{\rm{N,s}}$), i.e. how the magnetic spectrum is modified on the change of symmetry between low-temperature orthorhombic and high-temperature tetragonal phases. Recent resistivity measurements of de-twinned samples of Ba(Fe$_{1-x}$Co$_{x}$)$_{2}$As$_{2}$ show that there exists a large electronic anisotropy that persists above $T_{\rm{N,s}}$ \cite{Chu-Science}. Previous INS measurements of \CaAs above $T_{\rm{N,s}}$ \cite{Diallo above Tn} probed excitations only up to $\sim 60$\,meV. Given the possible role of magnetic fluctuations in the origin of the superconductivity of iron pnictides, further data on the magnetic spectrum in the tetragonal phase is of great interest.

The questions we set out to answer here are, firstly, are the magnetic interactions in \SrAs anisotropic in the ordered state, as in other 122-arsenides, secondly, do the spin excitation spectra change significantly on warming above $T_{\rm{N,s}}$, and thirdly, how robust is the local-moment description of the INS spectra? The INS results we describe here probe the magnetic excitations throughout the Brillouin zone (BZ) over the energy range $5<E<260$\,meV, below and above $T_{\rm{N,s}}$. We find that within a localized model the magnetic exchange parameters are strongly anisotropic below and above $T_{\rm{N,s}}$, and that an itinerant model gives the better qualitative description of the data. The form of the data is not significantly altered on warming above $T_{\rm{N,s}}$.

\section{Experimental Details}

Single crystals of \SrAs were grown by the self-flux method \cite{Gillett crystal growth,Sebastian-JPCM-2008}. The crystals are highly homogenous, as verified by microprobe analysis, a well-defined magnetic/structural transition temperature of $T_{\rm{N,s}} = 192$\,K, and by the observation of quantum oscillations from samples from the same growth \cite{Sebastian-JPCM-2008}.

\begin{figure}[!h]
\includegraphics*[scale=0.38,angle=0]{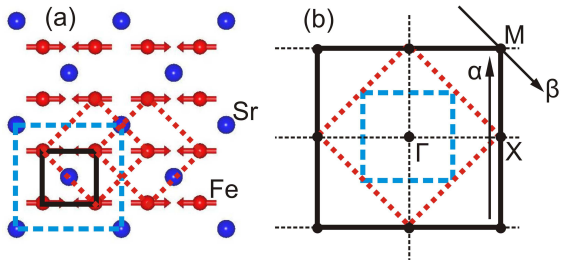}
\centering \caption{(color online). (a) Projection of the real space crystal structure onto the $ab$-plane, with As atoms omitted for clarity. The solid (black) square is the unit cell of the Fe sublattice, the dashed (blue) square is the orthorhombic unit cell (space group $Fmmm$), and the dotted (red) diamond is the tetragonal $I4/mmm$ unit cell. (b) Corresponding Brillouin zones. Referred to the Fe sublattice the standard square-lattice symmetry points are $\Gamma=(0,0)$, $\rm{X}=(0.5,0)$, and $\rm{M}=(0.5,0.5)$, and X is the AFM ordering wavevector. The arrows labeled $\alpha$ and $\beta$ indicate the direction of 1-d cuts and 2-d slices shown in  Fig.~\ref{fig:dogleg_slices} and Fig.~\ref{fig:k_diag_4panel}.}\label{fig:Fe_plane}
\end{figure}

On cooling through $T_{\rm{N,s}}$, the crystal symmetry changes from $I4/mmm$ to $Fmmm$, and an AFM structure develops with propagation vector $\mathbf{Q}_{\rm{AFM}}=(0.5,0.5,1)$ referred to the $I4/mmm$ unit cell [wave vectors are given in reciprocal lattice units (r.l.u.)]. As the magnetic dynamics in \SrAs are relatively two-dimensional we will henceforth give only in-plane wave vector components. Further, from now on we index wave vectors with respect to the square unit cell formed by the Fe atoms in the paramagnetic phase ($a_{\rm Fe} = b_{\rm Fe} = 2.8$\,{\AA}). In this convention $\mathbf{Q}_{\rm{AFM}}=(0.5,0)$. The unit cells and their corresponding Brillouin zones are shown in Fig.~\ref{fig:Fe_plane}.

The INS experiments were performed on the MERLIN time-of-flight (ToF) chopper spectrometer at the ISIS facility \cite{Merlin description}. Twenty one single crystals were co-aligned to give a mosaic sample of mass 5.4\,g, with a uniform mosaic of 4$^{\circ}$ (full-width at half-maximum). Spectra were recorded at temperatures of 6\,K, 212\,K ($T_{\rm{N,s}}+20$\,K), and 300\,K with neutrons of incident energy $E_{\rm i} = 50$, 100, 180, 300 and 450\,meV. The sample was aligned with the $c$ axis parallel to the incident neutron beam, and the $a$ axis horizontal. Data from equivalent positions in reciprocal space were averaged to improve statistics.  The scattering from a standard vanadium sample was used to normalize the spectra, $S(\mathbf{Q},E)$, and place them on an absolute intensity scale, with units mb\,sr$^{-1}$\,meV$^{-1}$\,f.u.$^{-1}$, where 1\,mb\,$=10^{-31}$\,m$^{2}$ and f.u. stands for `formula unit' of \SrAs.

\section{Results and Analysis}

\begin{figure}[ht]
\includegraphics*[scale=0.44,angle=0]{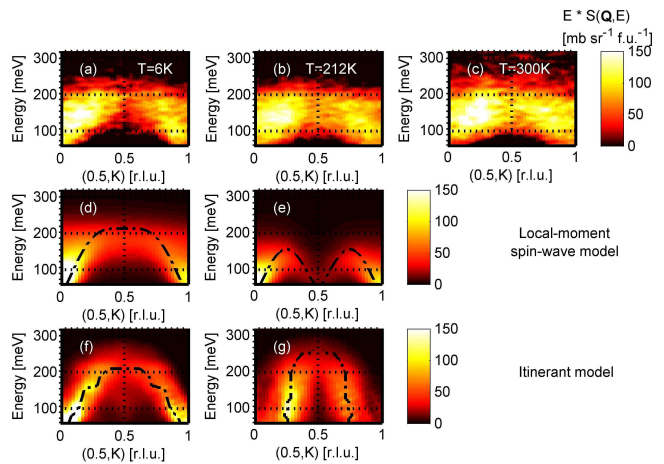}
\centering \caption{(color online). Energy--wavevector slices through the INS datasets at (a) $T=6$\,K, (b) $T=212$\,K, and (c) $T=300$\,K. Panels (d) and (e) show simulations of the spectrum over the same $\mathbf{Q}$ and energy range using a local-moment spin wave model with $J_{1a}\neq J_{1b}$, $J_{1b}<0$, and $J_{1a}=J_{1b}$ respectively. Panels (f) and (g) show the calculated $\chi''(\mathbf{Q},E)$ for the AFM phase, and the paramagnetic phase respectively; both taken from ref.~\onlinecite{Kaneshita-PRB}, and convoluted with the instrumental resolution, with the energy rescaled by a factor 0.85. Data are from the run with $E_{\rm i}=450$\,meV, and the intensity of both data and simulations have been multiplied by $E$ to improve clarity. The Fe form factor has been included in the simulations. Dashed lines in panels (d) and (e) indicate the simulated dispersion relations for the high energy parameters (see caption to Fig.~\ref{fig:k_diag_4panel}) and for $J_{1a}=J_{1b}$ respectively. Dashed lines in panels (f) and (g) are the loci of maximum intensity for the calculated $\chi''(\mathbf{Q},E)$. (The calculated low energy incommensurate behavior in (g) is discussed in ref.~\onlinecite{Kaneshita-PRB} and likely arises from limitations of the model. See text for further discussion of this point.)} \label{fig:dogleg_slices}
\end{figure}

\begin{figure}[h]
\includegraphics*[scale=0.45,angle=0]{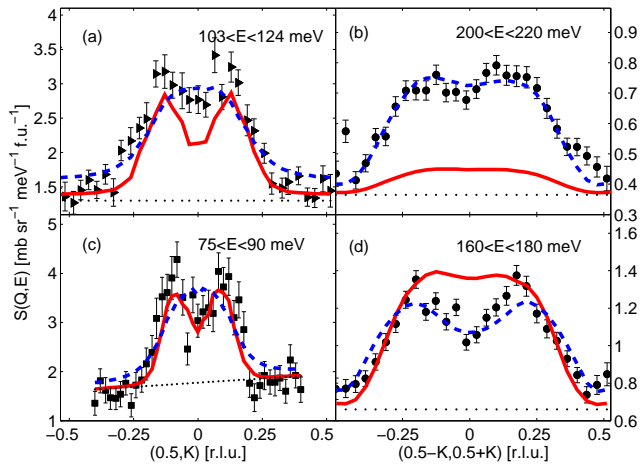}
\centering \caption{(color online). Wavevector cuts through the data taken at $T=6$\,K with incident neutron energies (c) $E_{\rm i}=180$\,meV; and (a), (b) and (d) $E_{\rm i}=450$\,meV. Solid (red) and dashed (blue) lines are calculated from fits to the local-moment $J_{1a}-J_{1b}-J_{2}$ model with the low-energy and high-energy best fit parameters, respectively. For lower energies best fits were obtained with $S_{\rm{eff}}=0.30\pm0.01$, $SJ_{1a}=30.8\pm1$\,meV, $SJ_{1b}=-5\pm5$\,meV, $SJ_{2}=21.7\pm0.4$\,meV, and $SJ_{c}=2.3\pm0.1$\,meV. This is in contrast to higher energies, where we found $S_{\rm{eff}}=0.69\pm0.02$, $SJ_{1a}=38.7\pm2$\,meV, $SJ_{1b}=-5\pm5$\,meV, and $SJ_{2}=27.3\pm0.7$\,meV. $SJ_{c}$ cannot be determined from cuts taken above the maximum of the dispersion along $(0,0,L)$ of $\sim 53$\,meV. Dotted (black) lines indicate the estimated non-magnetic scattering at $\mathbf{Q}=(1,0)$.}\label{fig:k_diag_4panel}
\end{figure}

The general form of the scattering at $T=6$\,K $\ll T_{\rm{N,s}}$ is illustrated by the energy--momentum slice in Fig.~\ref{fig:dogleg_slices}(a). The magnetic spectrum is similar to that of \CaAs \cite{Pengcheng Ca122,Diallo Ca122}, with intensity dispersing out of the $\mathbf{Q}_{\rm{AFM}}$ positions. The dispersion is revealed in more detail in Fig.~\ref{fig:k_diag_4panel}, which presents wave vector scans at four different energies. A single peak centered on $\mathbf{Q}_{\rm{AFM}}=(0.5,0)$ at low energies develops into a pair of peaks at $\sim75$\,meV which continue to separate and broaden with increasing energy. These peaks converge on $\mathbf{Q}=(0.5,0.5)$ at $E\sim 230$\,meV. Figs.~\ref{fig:dogleg_slices}(b) and (c) show that at  $T=212$\,K $>T_{\rm{N,s}}$ and $T=300$\,K $\gg T_{\rm{N,s}}$ the spectrum remains very similar to that of the ordered state, which means that strong AFM correlations persist well into the paramagnetic phase.

\begin{figure}[h]
\includegraphics*[scale=0.41,angle=0]{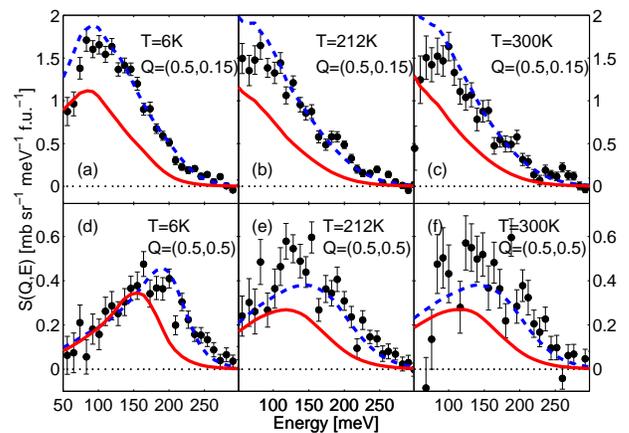}
\centering \caption{(color online). Energy cuts at constant in-plane $\mathbf{Q}$ through the data taken with $E_{\rm i} = 450$\,meV. The left-hand panels show data at $T=6$\,K, with (a) $\mathbf{Q}=(0.5,0.15)$  and (d) $\mathbf{Q}=(0.5,0.5)$. The middle panels (b) and (e) show data at $T=212$\,K at the same two wavevectors. The right-hand panels (c) and (f) show data at $T=300$\,K, also at the same two wavevectors. Solid (red) and dashed (blue) lines are calculated from the low-energy and high-energy fit parameters, respectively. The dotted (black) lines indicate the background, which is zero here because the non-magnetic background signal from $\mathbf{Q}=(1,0)$ has been subtracted from these data for clarity. The calculations for different temperatures differ only in the damping fitted -- the exchange parameters are the same for all six cuts.}\label{fig:energy_cut_4panel}
\end{figure}

Fig.~\ref{fig:energy_cut_4panel} provides a comparison of the spectra recorded at 6\,K, 212\,K and 300\,K for wave vectors near to $(0.5,0)$ and $(0.5,0.5)$. The signals at 212\,K and 300\,K are somewhat broader than that at 6\,K, but otherwise there are no marked differences associated with the change of crystal structure, even well above $T_{\rm{N,s}}$.

We first compare the data with the linear spin-wave spectrum calculated from a local-moment Heisenberg Hamiltonian. This model is the standard one used to interpret the spin excitations in the parent 122 compounds \cite{Our Ba122,Diallo Ca122,Pengcheng Ca122}, and is described in detail in the appendix. The key exchange parameters in this model are $J_{1a}$, $J_{1b}$ and $J_{2}$, which define the exchange along the $a$ and $b$ directions and the $ab$ diagonal respectively, as well as inter-planar exchange $J_{c}$

To fit data of the type presented we incorporated the resolution of the spectrometer in $(\mathbf{Q},E)$-space, since the widths of peaks in the scattering are often very strongly coupled to the instrumental resolution, which is itself a function of $E_{\rm i}$. We used the {\it Tobyfit} software \cite{Tobyfit_ref}, which uses Monte-Carlo methods to account for the effects of the $(\mathbf{Q},E)$ resolution of ToF spectrometers. We fitted the measured scattering cross-section, in the form of a set of one-dimensional cuts, over the entire energy range for which magnetic excitations were extant, convoluting the cross section with the spectrometer resolution.

Neutron inelastic scattering measurements of \SrAs performed on a triple-axis spectrometer with relatively low energy transfers \cite{Pengcheng Sr122 TAS} reveal a single gap of 6.5\,meV in the excitation spectrum. There is no evidence for two gaps, so we set the two modes described by equation (\ref{eq: dispersion}) in the appendix to be degenerate, i.e. $C=0$ in equations (\ref{eq: ACD defs}) and the in-plane and out-of-plane single-ion anisotropy terms are equal ($K_{ab}=K_{c}=K$). The single-ion anisotropy terms are thus determined by the size of the gap $\Delta$, which is given by

\begin{equation}
\frac{\Delta^2}{16 S^{2}} = K^2 + [J_{1a}+2J_{2}+J_{c}]K
\label{eq:aniso gap}
\end{equation}

Because the $c$ axis was parallel to the incident neutron beam, the $(0,0,L)$ component of the spectrum was coupled to time-of-flight, and hence to excitation energy. However, by using several $E_{\rm i}$ one can determine the dispersion along $(0,0,L)$, and hence determine the inter-layer exchange $J_{c}$. We found, as expected from the calculated response functions \cite{Our Ba122}, that at low temperature the cross-section is a maximum for odd-integer values of $L$, and minimum for even-integer values, and that $SJ_{c}=2.3\pm0.1$\,meV. For energies above the BZ boundary energy along $(0,0,L)$, for which $K=0$, $\{16(2SJ_{2}+SJ_{1a})SJ_{c}\}^{1/2} \simeq 53$\,meV, the INS spectrum is almost independent of $L$. Thus the fits performed where only data above $\sim 100$\,meV were considered could not constrain $J_{c}$, in which case we fixed $J_{c}$ to the value determined from the low-energy fits.

\section{Discussion}

Superficially, the spin-wave model provides a reasonable overall description of the low-temperature data, as illustrated in  Fig.~\ref{fig:dogleg_slices}(d). One robust outcome from the analysis was a large difference between $J_{1a}$ and $J_{1b}$ (see caption to Fig.~\ref{fig:k_diag_4panel}), similar to the anisotropy found in \CaAs \cite{Pengcheng Ca122}. Another clear finding was that the damping term in the spectrum is energy-dependent, with a steady increase at low energies followed by a rapid increase at about 80\,meV.

However, the spin-wave model fails in two important respects.  Firstly, good fits could only be achieved by fitting the lower energy ($E\lesssim 100$\,meV) and higher energy ($E\gtrsim 100$\,meV) parts of the low-temperature INS spectra separately -- the low-energy parameter set is unable to account for the existence of an appreciable signal above $\sim 200$\,meV, as seen in Fig.~\ref{fig:k_diag_4panel}(b) and Fig.~\ref{fig:energy_cut_4panel}(d), while the high-energy parameters predict that the spin-wave branches below $\sim 150$\,meV are unresolved, inconsistent with the data in Figs.~\ref{fig:k_diag_4panel}(a) and (c). Secondly, the high-temperature spectra are inconsistent with the tetragonal symmetry, which constrains $J_{1a}=J_{1b}$ for $T>T_{\rm{N,s}}$. This is illustrated in Fig.~\ref{fig:dogleg_slices}(e), which shows that the spin-wave spectrum when $J_{1a}=J_{1b}$ is gapless at $(0.5,0.5)$. The origin of this softening is frustration. When $J_{1a}=J_{1b}$, the nearest-neighbor interactions are frustrated. The AFM structure can then be regarded as two decoupled, interpenetrating AFMs, each on a square lattice of dimensions $\sqrt{2}a_{\rm Fe}\times \sqrt{2}a_{\rm Fe}$, the real space and Brillouin zone of which are shown as dotted (red) lines in Fig.~\ref{fig:Fe_plane}. The magnetic unit cell and corresponding magnetic Brillouin zone for the uncoupled AFMs are indicated by the dashed (blue) squares in Figs.~\ref{fig:Fe_plane}(a) and (b). The wavevectors $\rm{X}=(0.5,0)$, and $\rm{M}=(0.5,0.5)$ are both magnetic zone centers for the uncoupled AFMs, and therefore equivalent by symmetry. For the Fe sublattice, however, $\rm{X}$ and $\rm{M}$ are not equivalent by symmetry, and $\rm{M}$ is on the magnetic zone boundary for an itinerant AFM with $\mathbf{Q}_{\rm{AFM}}=(0.5,0)$. So the fact that the spin-wave energy goes to zero at $(0.5,0.5)$ for the tetragonal structure is purely a property of the local-moment treatment of the magnetic interactions.

These results raise the question as to whether there exists a mechanism that maintains an anisotropic exchange coupling ($J_{1a}\ne J_{1b}$) in the paramagnetic phase, or whether the local-moment model must be abandoned. One possible mechanism is electronic nematic order, which has been widely discussed in connection with anisotropic properties of several other classes of strongly-correlated electron systems \cite{Borzi-Science-Sr3Ru2O7, Lilly-PRL, Kee-PRB-nematic}. The existence of an electronic nematic state has recently been proposed to explain anisotropy in the in-plane resistivity \cite{Chu-Science} and elastic properties \cite{Fernandes-PRL-ElasticProps} of iron pnictides. In some cases the electronic nematic order is predicted to persist above $T_{\rm{N,s}}$ [ref.~\onlinecite{Fang-PRB77-Nematic}], although by only a few degrees at most. Thus, any evidence of broken symmetry above $T_{\rm{N,s}}$  is unlikely to be due to nematic order directly, but could perhaps be ascribed to nematic-like fluctuations, as in ref.~\cite{Fernandes-PRL-ElasticProps}, in which it is suggested that nematic fluctuations might persist up to room temperature. Our higher temperature measurements were performed at both 20\,K above $T_{\rm{N}}$ and room temperature, and as can be seen in Fig.~\ref{fig:dogleg_slices}(b) and (c), and Fig.~\ref{fig:energy_cut_4panel}, the neutron scattering spectra are very similar. Although one might expect some change in the spectra over such a wide temperature range, we cannot completely rule out the existence of nematic fluctuations.

Orbital ordering has also been suggested to explain the anisotropy \cite{Chen-PRB-OrbitalOrder}. In this case it has been predicted that a spin-wave mode would exist at high energies at $\mathbf{Q}=(0.5,0.5)$. The same calculations also indicate that on warming the energy of this mode softens slightly, and the peak height of the dynamical response function $S^{\alpha \beta}(\mathbf{Q},E)$ decreases rapidly. The broadening and lack of appreciable softening of the mode at $\mathbf{Q}=(0.5,0.5)$ agree well with our data. However, as shown in Fig.~\ref{fig:energy_cut_4panel}(c) to (e), the peak intensity is essentially unchanged, at variance with the calculation. Note that in ref.~\onlinecite{Chen-PRB-OrbitalOrder} the orbital order parameter is treated as an Ising-like parameter, so it is possible that if one considered instead a case in which there was partial orbital polarization the form of $S^{\alpha \beta}(\mathbf{Q},E)$ vs. temperature may be altered.

If the local-moment model cannot be reconciled with our data, as seems likely, then what about itinerant-electron models \cite{Eremin INS itinerant, Kaneshita-PRB}, or hybrid models that combine both local moments and itinerant electrons \cite{Lv-PRB}? In order to capture the key features of the data it is likely that a relatively detailed model will be required. Indeed, a comparison of the calculated $\chi''({\bf Q},E)$ from a minimal band model (ref.~\onlinecite{Eremin INS itinerant}) shows that it does not provide a good description of our data, except for $\lesssim 50$\,meV.

However, a mean-field model based on a more realistic 5-band structure \cite{Kaneshita-PRB} appears to give quantitatively good agreement with some of the features observed here. We illustrate this in Figs.~\ref{fig:dogleg_slices}(a) and (f), which show the low-temperature INS data together with the calculated dynamical susceptibility $\chi''(\mathbf{Q},E)$, convoluted with our instrumental resolution. We also over-plot the locus of maximum intensity from the itinerant model calculation. This curve is rather more structured than the smooth dispersion curves shown in Figs.~\ref{fig:dogleg_slices}(d) and (e), with several abrupt changes of gradient, which would of course be parameterized in a local-moment treatment by exchange parameters that changed with energy. The energy scale of the calculated $\chi''({\bf Q},E)$ has been changed by a factor $\sim0.85$ compared to the published calculation. This rescaling is likely due to the fact that in a mean-field approximation the energy scale typically needs to be renormalized down due to correlation effects not included in the model. The calculations in ref.~\onlinecite{Kaneshita-PRB} were performed with a Coulomb interaction $U=1.2$\,eV and Hund coupling $J=0.22$\,eV, chosen to yield the observed ordered moment of 0.8\,$\mu_{\rm{B}}$. There has, however, been some debate as to the strength of the electron correlations, as characterized by $U$ and $J$, in iron pnictides \cite{Haule-PRL-1111,Yang-PRB,Skornyakov-PRB}.

Fine tuning of the model parameters, i.e. $U$ and $J$, would no doubt improve the description of the data, but the present set seems to work reasonably well. Thus, the spin fluctuation spectrum of \SrAs that we measured indicates that this 122-arsenide exhibits rather weak electron correlations. In addition we note that the calculations from ref.~\onlinecite{Kaneshita-PRB} show that $\chi''(\mathbf{Q},E)$ does not soften at $\mathbf{Q}=(0.5,0.5)$ in the paramagnetic phase (Fig.~\ref{fig:dogleg_slices}(g)), which is also in agreement with our data (Fig.~\ref{fig:dogleg_slices}(b)). However, as noted in the caption of Fig.~\ref{fig:dogleg_slices}, the itinerant model of ref.~\onlinecite{Kaneshita-PRB} does have some shortcomings, in particular the form of the scattering at low energies in the paramagnetic phase. The calculation yields a signal at an incommensurate wavevector, which is not seen in the data. This incommensurability most likely arises from partial nesting in the paramagnetic band structure used in the mean-field calculation. Thus in order to obtain a better agreement between calculations and our results one would need to use a more detailed, experimentally determined, band structure, and also perhaps go beyond the mean-field approximation.

The key advantages of the itinerant model are thus that: (i) it results in a more structured signal, as observed, which the local-moment model can only explain with discontinuous exchange parameters; (ii) it gives an explanation, in the form of particle-hole excitations, for the energy-dependent damping; and (iii)  one does not need to invoke further phenomenology to explain the absence of a soft mode at $\mathbf{Q}=(0.5,0.5)$ in the paramagnetic phase.

As mentioned above, there have been proposed recently hybrid models that combine both local moments and itinerant electrons\cite{Lv-PRB}. In the AFM phase the calculations yield a high-energy mode at $\mathbf{Q}=(0.5,0.5)$, as we observe. They also yield a rather more structured dispersion relation than a straightforward local-moment treatment, which also qualitatively agrees with our measurements. It is not clear, however, how the spectra would change on warming above $T_{\rm{N,s}}$, which is a crucial discriminant between local-moment and itinerant models.

\section{Conclusions}

In conclusion, our analysis shows that although superficially a local-moment model can be used explain the nature of the spin fluctuations in \SrAs, close examination shows that it fails in several respects. In particular the data cannot be fitted with a single parameter set, and a soft mode does not appear at $\mathbf{Q}=(0.5,0.5)$ on warming above $T_{\rm{N,s}}$. On the other hand, an itinerant model appears to be able to explain both of these features. Thus it is not necessary to invoke further symmetry breaking, such as electronic nematic or orbital order, to explain the lack of soft mode at $\mathbf{Q}=(0.5,0.5)$.


\section{Acknowledgements}
We thank I. Eremin, Q. Si, and E. Kaneshita for helpful discussions. This work was supported by the Engineering and Physical Sciences Research Council, and the Science and Technology Facilities Council, of Great Britain.

\vspace{1mm}
\section*{Appendix: local-moment spin wave analysis}\label{a:lm_formalism}

The local-moment Hamiltonian used in the linear spin-wave analysis is:

\begin{equation}
H = \sum_{\langle jk\rangle}J_{jk}{\bf S}_j \cdot {\bf S}_k +
\sum_{j}\{K_c(S_z^2)_j + K_{ab}(S_y^2-S_x^2)_j\}.\label{eq: Hamiltonian}
\end{equation}

\noindent The first summation is over nearest-neighbor and
next-nearest-neighbor pairs with each pair counted only once. The
$J_{jk}$ are exchange parameters $J_{1a}$, $J_{1b}$ and $J_{2}$, along the $a$ and $b$ directions and the $ab$ diagonal respectively, as well as inter-planar exchange $J_{c}$. We also include in-plane and out-of-plane single-ion anisotropy constants $K_{ab}$ and $K_c$. Diagonalization of equation (\ref{eq: Hamiltonian}) leads to two non-degenerate branches with dispersion\cite{Our Ba122}

\begin{equation}
\hbar\omega_{1,2}({\bf Q}) = \sqrt{A_{\bf Q}^2-(C\pm D_{\bf
Q})^2},\label{eq: dispersion}
\end{equation}

\noindent where

\begin{eqnarray}
A_{\bf Q} & = & 2S\left \{J_{1b}[\cos({\bf Q}\cdot{\bf b})-1]+
J_{1a} + 2J_2+J_c\right\} \nonumber\\[2pt]
& & \hspace{80pt} + S(3K_{ab}+K_c)\nonumber\\[5pt]
C &= & S(K_{ab}-K_c)\nonumber\\[5pt]
D_{\bf Q} & = & 2S\{J_{1a}\cos({\bf Q}\cdot{\bf
a})+2J_2\cos({\bf Q}\cdot{\bf a})\cos({\bf
Q}\cdot{\bf b}) \nonumber\\[2pt]
& & \hspace{80pt} + J_c\cos(\frac{{\bf Q}\cdot{\bf c}}{2})\}.
\label{eq: ACD defs}
\end{eqnarray}

Note that in eq.~\ref{eq: ACD defs} {\bf a} and {\bf b} are the basis vectors of the Fe square lattice, whereas in Ref.~\onlinecite{Our Ba122} the corresponding equations are given with respect to the Fmmm space group. The in-plane cell parameters for the Fmmm space group are twice those of the Fe square lattice, as shown in Fig.~\ref{fig:Fe_plane}(a). Note also that in Ref.~\onlinecite{Our Ba122} a factor of two was missed from the term containing $J_{c}$ in the equation for $D_{\mathbf{Q}}$

The response functions per \SrAs formula unit for magnon creation, $S^{\alpha\beta}({\bf Q},\omega)$, which relate to the neutron scattering cross section \cite{Squires}, are given by\cite{Our Ba122}
\begin{eqnarray} S^{yy}({\bf Q},\omega) & =
& S_{\rm eff}\frac{A_{\bf Q}-C-D_{\bf Q}}{\hbar\omega_{1}({\bf Q})}
\{n(\omega)+1\}\delta[\omega-\omega_{1}({\bf Q})], \nonumber\\[5pt] S^{zz}({\bf Q},\omega) & =
& S_{\rm eff}\frac{A_{\bf Q}+C-D_{\bf Q}}{\hbar\omega_{2}({\bf Q})}
\{n(\omega)+1\}\delta[\omega-\omega_{2}({\bf Q})],\nonumber\\[5pt]\label{eq: response funcs}
\end{eqnarray}
\noindent where $S_{\rm eff}$ is the effective spin and $n(\omega)$ is the
boson occupation number. Only the transverse correlations ($yy$ and $zz$ for SrFe$_2$As$_2$)
contribute to the linear spin wave cross section. In linear spin wave theory $S_{\rm eff}=S$, but we keep them distinct here because in the analysis the natural independent parameters are the energies $SJ_{1a}, SJ_{1b}, ... SK_{ab}, SK_c$ and the effective spin $S_{\rm eff}$ of the fluctuating moment in the local moment approximation. To account for the finite lifetimes of the excitations we replace the delta-functions in equations (\ref{eq: response funcs}) by damped harmonic oscillator functions \cite{DSHO func}. Eqs.~(A2-4) were used to calculate the neutron scattering cross-section using Ref.~\onlinecite{Our Ba122}, Eq.~(4).

\end{document}